\documentclass[aps,prd,eqsecnum,showpacs,superscriptaddress,twocolumn,nofootinbib,
floatfix,preprintnumbers,altaffilletter]{revtex4}
\newif\ifpdf\ifx\pdfoutput\undefined\pdffalse\else\pdfoutput=1\pdftrue\fi
       \newcommand{\pdfgraphics}{\ifpdf\DeclareGraphicsExtensions{.pdf,.jpg}\else\fi}
\usepackage{graphicx}
\usepackage{hyperref}
\usepackage{amssymb}
\usepackage{amsmath}
\usepackage{longtable}
\usepackage{rotating}
\usepackage{color}
\usepackage{epsfig}
\usepackage{epsf}
\usepackage{fancyhdr}

\newcommand{\rd}{\,{\rm d}}
\def\bea{\begin{eqnarray}}
\def\eea{\end{eqnarray}}
\def\beq{\begin{equation}}
\def\eeq{\end{equation}}
\def\vec#1{\mathbf{#1}}

\newcommand{\ba}{\begin{eqnarray}}
\newcommand{\ea}{\end{eqnarray}}

\newcommand{\bml}{\begin{mathletters}}
\newcommand{\eml}{\end{mathletters}}

\def \calf {\cal F}

\def\bea{\begin{eqnarray}}
\def\eea{\end{eqnarray}}
\def\beq{\begin{equation}}
\def\eeq{\end{equation}}
\def\vec#1{\mathbf{#1}}

\begin{document}
\pdfgraphics
\pagestyle{fancy}

\title{
Setting upper limits on the strength of periodic gravitational waves
\\ using the first science data from the GEO\,600 and LIGO detectors\\ }
%

%
%
%
\newcommand*{\AG}{Albert Einstein Institut f\"ur Gravitationsphysik, D-14476 Golm, Germany}
\affiliation{\AG}
\newcommand*{\AH}{Albert Einstein Institut f\"ur Gravitationsphysik, D-30157 Hannover, Germany}
\affiliation{\AH}
\newcommand*{\AN}{Australian National University, Canberra, 0200, Australia}
\affiliation{\AN}
\newcommand*{\CH}{California Institute of Technology, Pasadena, CA  91125, USA}
\affiliation{\CH}
\newcommand*{\DO}{California State University Dominguez Hills, Carson, CA  90747, USA}
\affiliation{\DO}
\newcommand*{\CA}{Caltech-CaRT, Pasadena, CA  91125, USA}
\affiliation{\CA}
\newcommand*{\CU}{Cardiff University, Cardiff, CF2 3YB, United Kingdom}
\affiliation{\CU}
\newcommand*{\CL}{Carleton College, Northfield, MN  55057, USA}
\affiliation{\CL}
\newcommand*{\CO}{Cornell University, Ithaca, NY  14853, USA}
\affiliation{\CO}
\newcommand*{\FN}{Fermi National Accelerator Laboratory, Batavia, IL  60510, USA}
\affiliation{\FN}
\newcommand*{\HC}{Hobart and William Smith Colleges, Geneva, NY  14456, USA}
\affiliation{\HC}
\newcommand*{\IU}{Inter-University Centre for Astronomy  and Astrophysics, Pune - 411007, India}
\affiliation{\IU}
\newcommand*{\CT}{LIGO - California Institute of Technology, Pasadena, CA  91125, USA}
\affiliation{\CT}
\newcommand*{\LM}{LIGO - Massachusetts Institute of Technology, Cambridge, MA 02139, USA}
\affiliation{\LM}
\newcommand*{\LO}{LIGO Hanford Observatory, Richland, WA  99352, USA}
\affiliation{\LO}
\newcommand*{\LV}{LIGO Livingston Observatory, Livingston, LA  70754, USA}
\affiliation{\LV}
\newcommand*{\LU}{Louisiana State University, Baton Rouge, LA  70803, USA}
\affiliation{\LU}
\newcommand*{\LE}{Louisiana Tech University, Ruston, LA  71272, USA}
\affiliation{\LE}
\newcommand*{\LL}{Loyola University, New Orleans, LA 70118, USA}
\affiliation{\LL}
\newcommand*{\MP}{Max Planck Institut f\"ur Quantenoptik, D-85748, Garching, Germany}
\affiliation{\MP}
\newcommand*{\ND}{NASA/Goddard Space Flight Center, Greenbelt, MD  20771, USA}
\affiliation{\ND}
\newcommand*{\NA}{National Astronomical Observatory of Japan, Tokyo  181-8588, Japan}
\affiliation{\NA}
\newcommand*{\NO}{Northwestern University, Evanston, IL  60208, USA}
\affiliation{\NO}
\newcommand*{\SC}{Salish Kootenai College, Pablo, MT  59855, USA}
\affiliation{\SC}
\newcommand*{\SE}{Southeastern Louisiana University, Hammond, LA  70402, USA}
\affiliation{\SE}
\newcommand*{\SA}{Stanford University, Stanford, CA  94305, USA}
\affiliation{\SA}
\newcommand*{\SR}{Syracuse University, Syracuse, NY  13244, USA}
\affiliation{\SR}
\newcommand*{\PU}{The Pennsylvania State University, University Park, PA  16802, USA}
\affiliation{\PU}
\newcommand*{\TC}{The University of Texas at Brownsville and Texas Southmost College, Brownsville, TX  78520, USA}
\affiliation{\TC}
\newcommand*{\TR}{Trinity University, San Antonio, TX  78212, USA}
\affiliation{\TR}
\newcommand*{\HU}{Universit{\"a}t Hannover, D-30167 Hannover, Germany}
\affiliation{\HU}
\newcommand*{\BB}{Universitat de les Illes Balears, E-07071 Palma de Mallorca, Spain}
\affiliation{\BB}
\newcommand*{\BR}{University of Birmingham, Birmingham, B15 2TT, United Kingdom}
\affiliation{\BR}
\newcommand*{\FA}{University of Florida, Gainsville, FL  32611, USA}
\affiliation{\FA}
\newcommand*{\GU}{University of Glasgow, Glasgow, G12 8QQ, United Kingdom}
\affiliation{\GU}
\newcommand*{\MU}{University of Michigan, Ann Arbor, MI  48109, USA}
\affiliation{\MU}
\newcommand*{\OU}{University of Oregon, Eugene, OR  97403, USA}
\affiliation{\OU}
\newcommand*{\RO}{University of Rochester, Rochester, NY  14627, USA}
\affiliation{\RO}
\newcommand*{\UW}{University of Wisconsin-Milwaukee, Milwaukee, WI  53201, USA}
\affiliation{\UW}
\newcommand*{\WU}{Washington State University, Pullman, WA 99164, USA}
\affiliation{\WU}
\author{B.~Abbott}\affiliation{\CT}
\author{R.~Abbott}\affiliation{\CT}
\author{R.~Adhikari}\affiliation{\LM}
\author{B.~Allen}\affiliation{\UW}
\author{R.~Amin}\affiliation{\FA}
\author{S.~B.~Anderson}\affiliation{\CT}
\author{W.~G.~Anderson}\affiliation{\TC}
\author{M.~Araya}\affiliation{\CT}
\author{H.~Armandula}\affiliation{\CT}
\author{F.~Asiri}\altaffiliation[Currently at ]{Stanford Linear Accelerator Center}\affiliation{\CT}
% ...
 \collaboration{The LIGO Scientific Collaboration, http://www.ligo.org}
 \noaffiliation
\date{\today}
\begin{abstract}

Data collected by the GEO\,600 and LIGO interferometric
gravitational wave detectors during their first observational
science run were searched for continuous gravitational waves
from the pulsar J1939+2134 at twice its rotation frequency. Two
independent analysis methods were used and are demonstrated in
this paper: a frequency domain method and a time domain method.
Both achieve consistent null results, placing new upper limits on the
strength of the pulsar's gravitational wave emission. A
model emission mechanism is used to interpret the limits as a
constraint on the pulsar's equatorial ellipticity.

\end{abstract}
%
\maketitle
%
\section{INTRODUCTION}
\label{s:introduction}

This work presents methods to search for periodic gravitational
waves  generated by  known pulsars, using data collected by
interferometric gravitational wave detectors.  To illustrate these
methods, upper limits are placed on the strength of waves emitted
by pulsar J1939+2134 at its expected $1\,284$\,Hz emission
frequency during S1 \cite{S1_exp}. S1 is the first observational science 
run of the LIGO \cite{ligo1,ligo2} and GEO \cite{GEO1,GEO2} 
detectors and it took place during 17 days 
between August 23 and September 9, 2002.  
The sensitivity of the searches presented here surpasses that of previous searches for gravitational waves from this source. However, measurements of the
spin-down rate of the pulsar indicate that a detectable signal is
very unlikely given the instrument performance for this data set: 
for these early observations the detectors were not operating at
their eventual design sensitivities. Substantial improvements in
detector noise  have been achieved since the S1 observations, and
further improvements  are planned. We expect that the methods presented here 
will eventually enable the direct detection of periodic gravitational waves.

In Section~\ref{s:detectors}, we describe the configuration and
calibration of the four LIGO and GEO interferometers and derive their
expected sensitivities to periodic sources having known locations,
frequencies and spindown rates.  In Section~\ref{s:sources} we consider  proposed neutron
star gravitational wave emission mechanisms,  and introduce notation for
describing  the nearly monochromatic signals emitted by isolated neutron
stars.
Statistical properties of the data, analyses methods and results
are presented in Section~\ref{s:datanalysis}. These results are then summarized and
compared in Section~\ref{s:results}. In Section~\ref{s:results} we also interpret
the upper limits on the signal amplitude as a constraint on the
ellipticity of the pulsar and consider our results in the
context of previous upper limits.

\section{The Detectors}
\label{s:detectors} Gravitational waves are a fundamental
consequence of Einstein's General Theory of Relativity
\cite{einstein15, einstein16}, in which they represent
perturbations of the spacetime metric which propagate at the speed
of light.
Gravitational waves produced by the acceleration of compact
astrophysical objects may be detected by monitoring the motions
they induce on freely-falling test bodies. The strength of these
waves, called the \emph{strain,} can be characterized by the
fractional variation in the geodesic separation between these test
bodies.

During the past decade, several scientific collaborations have
constructed a new type of detector for gravitational waves. These
large-scale interferometric detectors include the US Laser
Interferometer Gravitational Wave Observatory (LIGO), located in
Hanford, WA, and Livingston, LA, built by a Caltech-MIT
collaboration \cite{ligo1,ligo2}; the GEO\,600 detector near
Hannover, Germany, built by a British-German collaboration
\cite{GEO1,GEO2}; the VIRGO detector in Pisa, Italy, built by an
Italian-French collaboration \cite{virgo97}; and the Japanese
TAMA\,300 detector in Tokyo \cite{tama95}.  In these detectors,
the relative positions of suspended test masses are sensed
interferometrically.  A gravitational wave produces a time-varying
differential displacement $\Delta L(t)$ in an interferometer that
is proportional to its arm length $L$.  The amplitude of the
gravitational wave is described by the dimensionless strain
$h(t)=\Delta L(t)/L$. For realistic periodic astrophysical sources
we typically expect strain amplitudes smaller than $10^{-24}$ .

The following sections introduce the operating configurations of
GEO\,600 and LIGO detectors during the S1 run. The references
provide more detailed descriptions of these detectors.

\subsection{Instrument configurations}
\label{ss:configuration} The GEO\,600 detector 
comprises a 4-beam Michelson delay line system of arm length
600\,m. The interferometer is illuminated by frequency-stabilized
light from an injection-locked Nd:YAG laser. Before reaching the
interferometer, the light is passed through two 8\,m triangular
mode-cleaning cavities.  During S1  approximately 2\,W of light was
incident on the interferometer.  A power recycling mirror of 1\%
transmission was installed to increase the effective laser power
available for the measurement.

LIGO comprises three
power-recycled Michelson interferometers with resonant Fabry-Perot
cavity arms. A 4\,km and a 2\,km interferometer are collocated at
the Hanford site and are designated H1 and H2
respectively, and a 4\,km interferometer at the Livingston
site is designated L1. Each interferometer employs a
Nd:YAG laser stabilized using a monolithic reference cavity and a 
12\,m mode-cleaning cavity. 

In all four instruments the beam splitters, recycling mirrors and
test masses are hung as pendulums from multilayer seismic
isolation filters to isolate them from local forces. The masses
and beam paths are housed in high vacuum enclosures to preclude
optical scintillation and acoustic interference.

Sinusoidal calibration forces of known amplitude
were applied to the test bodies throughout the observing run. These
signals were recovered from the data stream and used to periodically
update the scale factors linking recorded signal amplitude to strain.
The principal calibration uncertainties arise from the imprecision
in the electromechanical coupling coefficients of the force
actuators. These were estimated by comparison with the known laser wavelength,
by actuating a test mass between interference fringes. In the Hanford 
interferometers, the calibration was also verified against piezoelectric
displacement transducers connected to the mirror support structures.
For the S1 observations, the net amplitude uncertainty was estimated at
$\pm$4\% for GEO, $\pm$10\% for each of the LIGO interferometers.

\subsection{Expected sensitivity}
\label{ss:sensitivity}
We define the gravitational wave strength $h_0$ of a continuous
signal from a given source as the maximum peak 
amplitude which could be received by an interferometer if the 
orientation of the pulsar and the detector were both optimal. 
Thus, $h_0$ depends on the intrinsic
emission strength and on the source distance, but not on the
inclination of the pulsar's spin axis or on the antenna pattern of
the detector.

The calibrated interferometer strain output is a time series
\begin{equation}
 s(t) = h(t) + n(t),
\end{equation}
where $h(t)$ is the received signal, $n(t)$ is the detector noise
and $t$ is the time in the detector's frame.

The noise $n(t)$ is characterized by its single-sided power
spectral density $S_n(f)$. Assuming this noise is Gaussian and
taking some fixed observation time\footnote{ Here we presume that
we know the position, frequency and spin-down parameters of the
source and that $T$ is between a few days and several months. }
$T$, we can compute the amplitude $h_0$ of a putative continuous
signal which would be detected in, e.g., 90\% of experimental
trials if truly present, but would arise randomly from the noise
background in only 1\% of trials  (what we call a 1\% ``false
alarm rate" and  a 10\% ``false dismissal rate").

\begin{figure}[t]
\centering
\includegraphics[width=8cm]{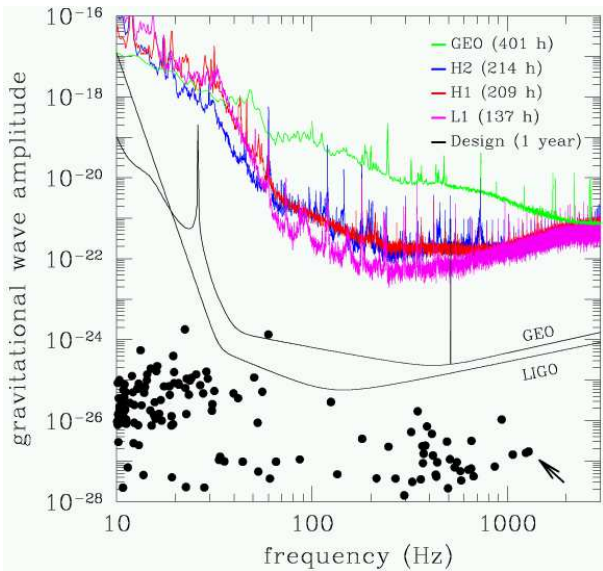}
\caption{Upper curves: characteristic amplitude $\langle h_0
\rangle $ of a known monochromatic signal detectable with a $1\%$
false alarm rate and a $10\%$ false dismissal rate by the GEO and
LIGO detectors at the S1 sensitivity and with an observation time
equal to the up-time of the detectors during S1 (GEO:\,401\,h,
L1:\,137\,h, H1:\,209\,h, H2:\,214\,h). Lower curves: $\langle h_0
\rangle$ for the design sensitivities of the detectors for an
assumed 1\,y observation time. Filled circles: upper limit on
$\langle h_0 \rangle$ from measured spin-down rate of known radio
pulsars assuming a moment of inertia of $10^{45}{\rm g~cm^2}$. These upper limits were derived under the assumption that all the measured loss of angular momentum of the star is due to the emission of gravitational waves, {\it{neglecting spin-down contribution from electromagnetic and particle emission}}. The
arrow points to the filled circle representing pulsar J1939+2134.}

\label{f:uls}
\end{figure}

If we fix a false alarm rate, it is clear that the lower the
desired false dismissal rate the higher the signal needs to be.
The detection statistic used in section \ref{ss:frequency}
provides the lowest false dismissal rate for a given false alarm
rate and signal strength and it is thus optimal in the
Neyman-Pearson sense (see for example \cite{allen02}). The
amplitude of the average signal that we could detect in Gaussian
stationary noise with a false alarm rate of 1\% and a false
dismissal rate of 10\% using the detection statistic
described in \cite{schutz2} is given by\footnote{The average is
over different positions, inclinations and polarizations of the
source.}
 \beq
 \label{uls}
 \langle h_0 \rangle = 11.4 \sqrt{S_n(f_{\rm s})/T},
 \eeq
where $f_{\rm s}$ is the frequency of the signal\footnote{ This
differs from \cite{bccs} for three reasons: 1) the $h_0$ used here
is twice that defined in \cite{bccs}, 2) we use a
different statistic for this detection problem (a chi-square
distribution with 4 degrees of freedom) and 3) we have specified a
false dismissal rate of 10\% whereas the derivation in \cite{bccs}
has an implicit false dismissal rate of about 50\%. If we use this
false dismissal rate and the $\calf$ statistic we get
$\langle h_0 \rangle =7.6 \sqrt{S_n(f_{\rm s})/T}$. }. The upper
curves in Fig.~\ref{f:uls} show $\langle h_0 \rangle$ for the LIGO
and GEO detectors during S1. Observation times for
respective interferometers are given in the figure. Because of ground motion, equipment failures and alignment drifts, the four interferometers were not always fully operational during the S1 run, thus the observation times vary from detector to detector.

The lower curves in Fig.~\ref{f:uls} represent $\langle
h_0\rangle$ corresponding to the design sensitivity of the various
detectors. An observation of $T=$ 1\,y was assumed.

The filled circles in Fig.~\ref{f:uls} show the constraints that
measurements of spin-down rates of known pulsars place on
the expected gravitational wave signal, under the assumption that
the pulsars are rigid rotators with a moment of inertia of
$10^{45}{\rm g\,cm^2}$ and {\it {that all the observed spin-down rate is
due to the emission of gravitational waves}}.

As shown in Fig.~\ref{f:uls}, under the above assumptions no detection is
expected for any known pulsar at the sensitivity
achieved during the S1 run. Furthermore, many known pulsars are rotating too
slowly to be detected by the initial ground-based interferometers. However,
the number of millisecond pulsars observed in this
band continues to increase with new radio surveys, and
the known targets plotted here constitute
a highly selected sample. Future searches for previously undiscovered
rotating neutron stars using the methods presented here will sample a different and potentially  much larger subset of the total population.

\section{Periodic gravitational waves}
\label{s:sources}

\subsection{Expected emission by neutron stars}
\label{s:s:sources}

The strongest argument that some neutron stars (NSs) {\it are}
emitting gravitational waves (GWs) with amplitude detectable by
Advanced LIGO ~\cite{AdLIGOproposal}, $h_0 \agt 10^{-27}-10^{-26}$, is due to
Bildsten~\cite{Bildsten98,Bildsten02}. He noted that the
inferred rotation frequencies of low-mass X-ray binaries (LMXBs)
are all clustered in the range $f_{\rm r} \sim
270-620$\,Hz (an inference strengthed by the recent observations of ~\cite{chakrabarty03,Wijnands03}), whereas \emph{a priori} there should be no cut-off
in $f_{\rm r}$, up to the (estimated) NS break-up frequency of
$\sim 1.5$\,kHz. Updating a suggestion by Wagoner~\cite{Wagoner84,Wagoner03},
Bildsten proposed that LMXBs have reached an equilibrium where
spin-up due to accretion is balanced by spin-down from GW
emission. Since the GW spin-down torque scales like $f_{\rm r}^5$,
a wide range of accretion rates then leads to a rather narrow
range of equilibrium rotation rates, as observed.

Millisecond pulsars (MSPs) are generally believed to be recycled
pulsars: old pulsars that were spun up by accretion during an LMXB
phase (\cite{verbunt93,klis00}). The rotation rates of MSPs also show a high-frequency 
cut-off~\cite{Bildsten02}; the fastest (PSR J1939+2134) has
$f_{\rm r} = 642$\,Hz. If the GWs that arrest the spin-up of
accreting NSs continue to be emitted in the MSP phase (e.g.,
because of some persistent deformation of the NS shape away from
axisymmetry), then they could also account for the observed
spin-down of MSPs. In this case, the GW amplitudes of MSPs would
in fact be (very close to) the `spin-down upper limits' shown in
Fig.~\ref{f:uls}. (Note that the MSP's spin-down rate is generally
attributed entirely to the pulsar's magnetic field; indeed, 
pulsar magnetic fields are typically inferred this way. However, there
appears to be no strong evidence supporting this inference.)

We now turn to the possible physical mechanisms responsible for
periodic GWs in this frequency range. The main possibilities that have been considered are
1) NS spin precession, 2) an excited NS oscillation mode (mostly likely
the r-mode), and 3) small distortions of the NS shape away from
axisymmetry. At present the third mechanism (small ellipticity)
seems the most plausible source of detectable GWs, and in this
paper we set upper limits for this particular mechanism (the three
mechanisms predict three different GW frequencies for the same
observed rotation frequency).  However, we begin by briefly
commenting on the other two possibilities.

A NS precesses (or `wobbles') when its angular momentum $\vec J$ is
not aligned with any principal axis of its inertia tensor. A wobbling
NS emits GWs at the inertial-frame precession frequency, which is very
nearly the rotation frequency, $f_{\rm r}$. While large-amplitude wobble could
plausibly produce GW amplitudes
$h_0 \sim 10^{-27}$
over short timescales, the
problem with this mechanism is that dissipation should damp NS
wobble quickly~\cite{DIJones}; while this dissipation timescale is
quite uncertain (it is perhaps of order a year for a MSP), it is
almost certainly orders of magnitude shorter than typical lifetimes of
MSPs. So unless some mechanism is found that regularly re-excites
large-amplitude wobble, it is unlikely that any nearby MSP would be
wobbling. Moreover, most MSPs have highly stable pulse shapes,
and typically appear {\it not} to be wobbling substantially.
In particular, the single-pulse characteristics of PSR J1939+2134 have
been observed to be extremely stable with no pulse-to-pulse variation
except for occasional giant pulses~\cite{jenet_01}.
It has been verified through radio observations that PSR J1939+2134
continued to spin according to a simple spin-down model during S1 \cite{lommen03}.

R-modes (modes driven by Coriolis forces) have been a source of
excitement among GW theorists since 1998, when
Andersson~\cite{andersson_98} and Friedman and Morsink~\cite{friedman_98}
showed that they should be unstable
due to gravitational back-reaction (the Chandrasekhar-Friedman-Schutz
instability).  Nonlinear mode-mode coupling is predicted to saturate
the growth of r-modes at dimensionless amplitude $\alpha \alt
10^{-3}(f_{\rm r}/{\rm kHz})^{5/2}$~\cite{arras_2002}.  This implies
r-mode radiation from nascent NSs in extragalactic supernovae will not be
detectable, but r-mode GWs from old, recycled Galactic NSs
could still be detectable
by Advanced LIGO.  For example, GWs from an excited r-mode could balance
accretion torque in accreting NSs, as in the Wagoner-Bildsten
mechanism.

We now turn to GWs from small non-axisymmetries in the NS shape.  If
$h_0$ is the amplitude of the signal at the detector from an optimally oriented source, as described above, and if we assume that the emission mechanism is due to
deviations of the pulsar's shape from perfect axial symmetry, then
\begin{equation}\label{h0_eps}
h_0 = \frac{4\pi^2 G}{c^4}\frac{I_{zz}f_{\rm s}^2}{r}\epsilon,
\end{equation}
where $r$ is the distance to the NS, $I_{zz}$ is its principal moment
of inertia about the rotation
axis, $\epsilon \equiv (I_{xx}-I_{yy})/I_{zz}$ is its
ellipticity, and the gravitational wave signal frequency, $f_{\rm s}$, is exactly
twice the rotation frequency, $f_{\rm r}$. $G$ is Newton's constant, and $c$ is 
the speed of light.  This is the emission mechanism that we
assume produces the gravitational wave signal that we are targeting.

One possible source of ellipticity is tiny `hills' in the NS
crust, which are supported by crustal shear stresses. In this
case, the maximum ellipticity is set by the crustal breaking
strain $\bar\sigma_{\rm max}$~\cite{UCB00}:
\begin{equation}\label{eps_max}
\epsilon_{\rm max} \approx 5 \times 10^{-8}
\bigl(\bar\sigma_{\rm max}/10^{-3}\bigr) \, .
\end{equation}
The coefficient in Eq.~\ref{eps_max} is low both because a NS's
crust is rather thin (compared to the NS radius), and because the
crust's shear modulus $\mu$ is small compared to the ambient pressure $p$:
$\mu/p \sim 10^{-3}-10^{-2}$.  (If NSs have solid cores, as well
as crusts, then much larger ellipticities would be possible.)  For
the LMXBs, Ushomirsky, Cutler \& Bildsten~\cite{UCB00} showed that
lateral temperature variations in the crust of order $5 \%$, or
lateral composition variations of order $0.5 \%$ (in the
charge-to-mass ratio), could build up NS ellipticities of order
$10^{-8}-10^{-7}$, but only if the crust's breaking strain is
large enough to sustain such hills.

Strong internal magnetic fields are another possible source of NS
ellipticity. Cutler~\cite{Cutler02} has argued that if a NS's
interior magnetic field $B$ has a toroidal topology (as expected if the $B$
field was generated by strong differential rotation immediately
after collapse), then dissipation tends to re-orient the symmetry
axis of the toroidal $B$ field perpendicular to the rotation axis,
which is the ideal orientation for maximizing equatorial
ellipticity. Toroidal $B$ fields of order $10^{12}-10^{13}$\,G
would lead to sufficient GW emission to halt the spin-up of LMXBs
and account for the observed spin-down of MSPs.

\subsection{The signal received from an isolated pulsar}
\label{s:signalmodel}
A gravitational wave signal we detect from an isolated pulsar will
be amplitude-modulated by the varying sensitivity of the detector
as it rotates with the Earth (the detector's `antenna pattern'). The 
detected strain has the form \cite{schutz2}
\begin{eqnarray}
h(t) &=& F_{+}(t,\psi)~h_{0}~\frac{1 + \cos^{2}\iota}{2}~\cos
\Phi(t)
\nonumber
\\
&+&  F_{\times} (t,\psi)~h_{0}~\cos \iota ~\sin  \Phi(t),
\label{h}
\end{eqnarray}
where $\iota$ is the angle between neutron star's spin direction
$\hat{\vec{s}}$ and the propagation direction of
the waves $\hat{\vec{k}}$ , and $\Phi(t)$ is the phase evolution of the signal.
$F_{+,\times}$ are the strain antenna patterns of the detector to
the plus and cross polarizations and are bounded between -1 and 1.
They depend on the 
orientation of the detector and the source, and on the polarization of the waves, described by the polarization angle $\psi$
\footnote{
Following the conventions of ~\cite{schutz2}, $\psi$ is the angle
(clockwise about $\hat{\vec{k}}$) from $\hat{\vec{z}} \times \hat
{\vec{k}}$ to $\hat{\vec{k}} \times \hat{\vec{s}}$, where $\hat
{\vec{z}}$ is directed to the North Celestial Pole.
$\hat{\vec{k}} \times \hat{\vec{s}}$ is the $x$-axis of the wave
frame -- also called the waves' principal $+$ polarization
direction.
}.

The signal will also be Doppler shifted by the orbital motion and
rotation of the Earth. The resulting phase evolution of the
received signal can be described by a truncated Taylor series as
\begin{eqnarray}
 \Phi(t) &=& \phi_{0} + 2\pi [f_{\rm s}(T - T_{0})
\nonumber
\\
&+&\frac{1}{2}\dot{f_{\rm s}}(T-T_{0})^{2} + \frac{1}{6}\ddot{f_{\rm s}}(T - T_{0})^{3}],
\label{phase1}
\end{eqnarray}
where 
\begin{equation}
T = t + \delta t= t - \frac{\vec{r_d} \cdot \hat{\vec{k}}}{c} +
\Delta_{E\odot} - \Delta_{S\odot}. \label{time}
\end{equation}
Here, $T$ is the time of arrival of a signal at the solar system 
barycentre (SSB), $\phi_{0}$ is the phase of the signal at 
fiducial time $T_{0}$, $\vec{r_d}$ is the position of the detector
with respect to the SSB, and $\Delta_{E\odot}$ and
$\Delta_{S\odot}$ are the solar system Einstein and Shapiro time
delays respectively \cite{T92}.

\begin{figure} \centering \includegraphics[height=5.5cm]{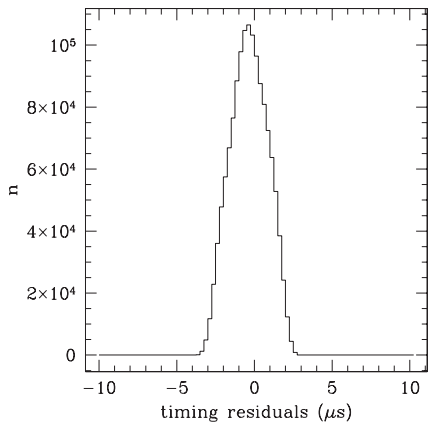}
\caption{Histogram of timing residuals between our barycentring
routines and TEMPO, derived by comparing the phase evolution of test signals produced by the two software packages.  156 locations in the sky were chosen at random
and the residuals calculated once an hour for the entire year 2002.  The
maximum timing error is $< 4\,\mu$s.}
\label{f:timingAccuracy}
\end{figure}

The timing routines used to compute the conversion between
terrestrial and SSB time (Eq. \ref{time}) were checked by comparison with the
widely-used radio astronomy timing package TEMPO \cite{tempo}.
This comparison (Figure~\ref{f:timingAccuracy}) confirmed an
accuracy of better than $\pm$ 4 $\mu$s, thus ensuring no more than
0.01 radian phase mismatch between a putative signal and its
template. This results in a negligible fractional signal-to-noise
ratio loss, of order $\sim 10^{-4}$.

Table \ref{t:pulsar_parameters} shows the parameters of the pulsar that
we have chosen to illustrate our analysis methods \cite{kaspi94}.
\begin{table}
\begin{center}
\begin{tabular}{rl}
  \hline
  right ascension (J2000) & $19^{\rm h}\,39^{\rm m}\,38^{\rm s}.560\,210(2)$ \\
  declination (J2000) & $+21^\circ\, 34^{\rm m}\, 59^{\rm s}.141\,66(6)$ \\
  RA proper motion & $-0.130(8)$ mas\,yr$^{-1}$ \\
  dec proper motion & $-0.464(9)$ mas\,yr$^{-1}$  \\
  period ($1/f_{\rm r}$) & $0.001\,557\,806\,468\,819\,794(2)\,{\rm s}$ \\
  period derivative & $1.051\,193(2)\times10^{-19}\,\rm{s}\,\rm{s}^{-1}$ \\
  epoch of period and position & MJDN 47\,500\\
  \hline
\end{tabular}
\end{center}
\caption{Parameters for the target pulsar of the analyses presented here, PSR J1939+2134 (also
designated PSR B1937+21). Numbers in parentheses indicate uncertainty
in the last digit.} \label{t:pulsar_parameters}
\end{table}

%

\section{Data Analyses}
\label{s:datanalysis}
\subsection{Introduction}
\label{ss:dataanalysisintro}

Two independent search methods are presented here: i) a frequency
domain method which can be employed for exploring large parameter
space volumes and ii) a time domain method for targeted searches
of systems with an arbitrary but known phase evolution. 

Both approaches will be used to cast an upper
limit on the amplitude of the periodic gravitational wave signal:
a Bayesian approach for the time domain analysis and a frequentist
approach for the frequency domain analysis.  These approaches provide
answers to two different questions and therefore should not be expected to result in the exact same numerical answer 
\cite{OHagan94,finn98}. The frequentist upper limit refers to the reliability of 
a procedure for identifying an interval that contains the true value of $h_0$. 
In particular, the frequentist confidence level is the fraction of putative observations in which, in 
the presence of a signal at the level of the upper limit value identified by the actual measurement, $h_0^{95\%}$, the 
upper limit identified by the frequentist procedure would have been higher than $h_0^{95\%}$. 
The Bayesian upper limit, on the other hand, defines an interval in $h_0$ that, based on the observation made and on 
prior beliefs, includes the true value with $95\%$ probability. The probability that we 
associate with the Bayesian upper limit characterizes the uncertainty in $h_0$ given the observation made. This 
is distinct from the reliability, evaluated over an ensemble of observations, of a procedure for identifying intervals. 

All the software used for the analyses is part of the
publicly available LSC Algorithm Library (LAL, \cite{lal}).  This is a 
library that comprises roughly 700 functions specific to gravitational
wave data analysis.

\subsection{Statistical characterization of the data}
\label{ss:characterization}

Due to the narrow frequency band in which the target signal has appreciable energy, it is most convenient to characterize the 
noise in the frequency domain. We divided the data into $60\,s$ blocks 
and took the Fourier transform of each. The resulting set of Fourier 
transforms will be referred to as SFTs (Short time-baseline Fourier 
Transforms) and is described in more detail in \ref{sss:sfts}.

\begin{figure} \centering \includegraphics[width=9.0cm]{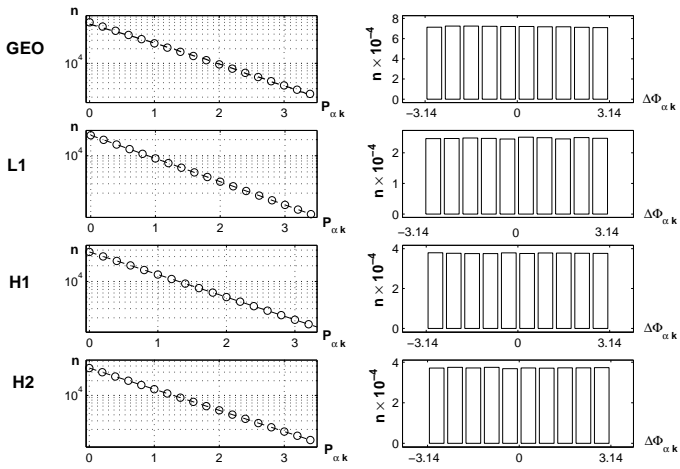}
\caption{Histograms of $P_{\alpha k}$ and of $\Delta \Phi_{\alpha k}$ for the four detectors.}
\label{f:SFTstatJ1939}
\end{figure}

The frequency of the pulsar signal at the beginning of the observation for
every detector is reported in
Table~\ref{t:freqSearchParams}. Also reported is the value of the spin-down
parameter expressed in units of Hz\,s$^{-1}$.
\begin{table}
\begin{center}
\begin{tabular}{ll}
  \hline
  spin-down & $-8.663\,3(43)\times10^{-14}$\,Hz\,s$^{-1}$\\
  $f_{\rm s}$ at start of GEO observation & $1\,283.856\,487\,705(5)$\,Hz\\
  $f_{\rm s}$ at start of L1 observation & $1\,283.856\,487\,692(5)$\,Hz\\
  $f_{\rm s}$ at start of H1 observation & $1\,283.856\,487\,687(5)$\,Hz\\
  $f_{\rm s}$ at start of H2 observation & $1\,283.856\,487\,682(5)$\,Hz\\
  \hline
\end{tabular}
\end{center}
\caption{Run parameters for PSR J1939+2134. The
different emission frequencies correspond to the different initial
epochs at which each of the searches began. Numbers in parentheses indicate the uncertainty
in the last digit or digits. }
\label{t:freqSearchParams}
\end{table}
We have studied the statistical properties of the data in a narrow
frequency band (0.5\,Hz) containing the emission frequency.
This is the frequency search region, as well as the region used for
estimating both the noise background and the detection efficiency.
Fig.~\ref{f:SFTstatJ1939} summarizes our findings. Two types of
distributions are plotted.  The first column shows the
distributions of bin power; for each SFT (labelled by $\alpha$)
and for every frequency bin (labelled by $1 \leq k \leq M$) in the band
1\,283.75 to 1\,284.25\,Hz, we have computed the quantity \beq
P_{\alpha k} = \frac{|\tilde x_{\alpha k}|^2}{\sum^{M}_{k} |\tilde
x_{\alpha k}|^2/M}, \eeq where $\tilde x_{\alpha k}$ is the SFT
datum at frequency index $k$ of the $\alpha$-th SFT, and have
histogrammed these values.  If the data are Gaussian and if the
different frequency bins in every SFT are independent realizations
of the same random process, then we expect the normalized power
variable described above ($P_{\alpha k}$) to follow an exponential
distribution with a mean and standard deviation of 1, as shown by
the dashed line. The circles are the experimental points.
The standard deviation of the measured distribution for GEO data is $0.95$. 
The LIGO Livingston, Hanford
4\,km and Hanford 2\,km data are also shown in
Fig.~\ref{f:SFTstatJ1939}. The standard deviation of the $P_{\alpha k}$
for all of these is $0.97$.

The plots in the second column of Fig.~\ref{f:SFTstatJ1939} show
the distribution of phase differences between adjacent frequency
bins. With the same notation as above, we have computed the
quantity \beq \Delta \Phi_{\alpha k} = \Phi_{\alpha k} -
\Phi_{\alpha k-1}, \eeq where $\Phi_{\alpha k}$ is the phase of
the SFT datum at frequency index $k$ of the $\alpha$-th SFT and
the difference is reduced to the range $[-\pi,\pi]$. $ \Delta
\Phi_{\alpha k}$ is therefore the distance in phase between data
at adjacent frequency bins. If the data were from a  purely random
process we expect this distribution to be uniform between $-\pi$
and $\pi$, as observed.

\begin{figure} \centering \includegraphics[width=9.0cm]{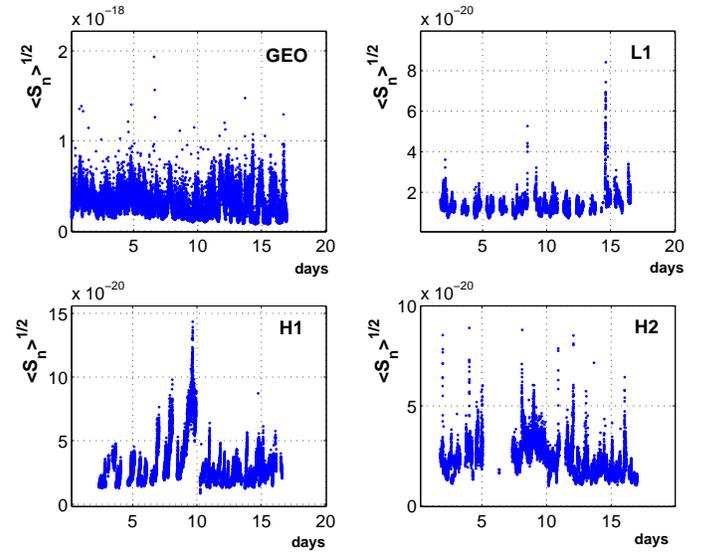}
\caption{The square root of the average value of $S_n$ for all four interferometers
over a band of 1\,Hz starting at 1\,283.5\,Hz versus time in days starting at the
beginning of S1 (August 23 2002, 15:00 UTC).}
\label{f:PsdAvgVsTimeJ1939}
\end{figure}

Fig.~\ref{f:PsdAvgVsTimeJ1939}
shows the average value of
$\sqrt{S_n}$ over a 1\,Hz band from 1\,283.5 to 1\,284.5\,Hz
as a function of time in days for the entire S1 run starting from the
beginning of S1 (15:00 UTC, August 23 2002).
These plots monitor the stationarity
of the noise in the band of interest over the course of the run.

\begin{figure} \centering \includegraphics[width=9.0cm]{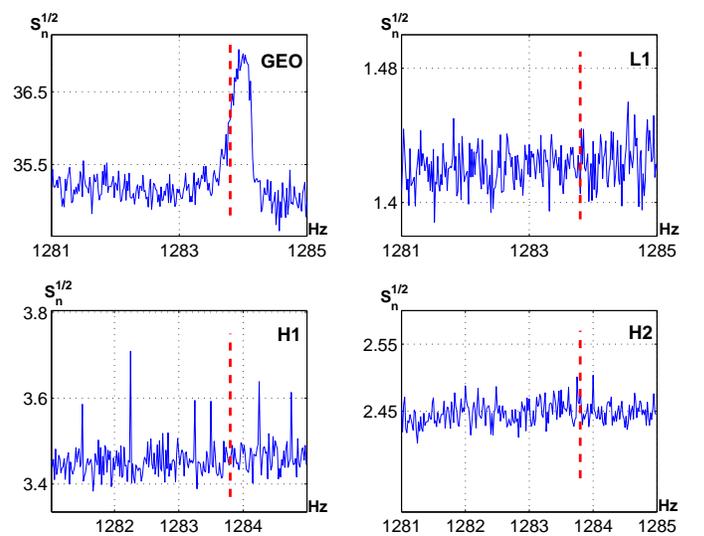}
\caption{$\sqrt{S_n}$ in a band of 4\,Hz
(starting at 1\,281\,Hz)  using the entire S1 data set
analyzed from the four interferometers. The noise $\sqrt{S_n}$ is shown in units
of $10^{-20}$\,Hz$^{-1/2}$.  The dashed vertical line indicates the expected
frequency of the signal received from J1939+2134.}
\label{f:S1psdj1939}
\end{figure}

Fig.~\ref{f:S1psdj1939} shows $\sqrt{S_n}$ as a function of
frequency between 1\,281 and 1\,285\,Hz.  During S1 the received
signal is expected to have a frequency 1\,283.8 Hz. This frequency
is shown as a dashed vertical line. During the S1 observation time
the Doppler modulation changed this signal frequency by no more
than $~ 0.03$ Hz, two SFT frequency bins. For these plots $S_n$
has been estimated by averaging the power in each frequency bin
over the entire S1 run. A broad spectral feature is observed in
the GEO data. This feature is $~ 0.5$ Hz wide, comparatively broad
with respect to the expected Doppler shift of the target signal,
and represents only a 10\% perturbation in the local power
spectral density.
\subsection{The frequency domain technique}
\label{ss:frequency}

\subsubsection{The Short time baseline Fourier Transforms}
\label{sss:sfts}
In principle the only constraint on the time baseline of the SFTs
used in the frequency domain analysis is that the instantaneous
frequency of a putative signal does not shift during the time
baseline by more than half a frequency bin. For frequencies in the
kiloHertz range this implies a maximum time baseline of order
30\,min (having assumed an observation time of several months and
a source declination roughly the same as the latitude of the
detector). However, in practice, since we are also estimating the
noise on the same time baseline, it is advisable for the time
baseline to be short enough to follow the non-stationarities of
the system. On the other hand, for the frequency-domain analysis,
the computational time required to carry out a search increases
linearly with the number of Fourier transforms. Thus the shorter
the time baseline, the higher the computational load. We have
chosen for the S1 run a time baseline of 60\,s as a compromise
between the two opposing needs.

Interruptions in interferometer operation  broke each time series
into segments separated by gaps representing invalid  or contaminated data.
Only valid data segments were included in the analysis. Each valid
60\,s data segment was filtered with a fifth-order Butterworth
high-pass filter having a knee frequency of 100\,Hz. Then, a
nearly flat-top Tukey window function was applied to each data
segment in the time domain.
The window changes the value of less than 1\% of the data in each
60\,s chunk. Each data segment was then Fast Fourier Transformed
and written to an SFT file.  These SFTs were computed once and
then used repeatedly for different analyses.

\subsubsection{The $\calf$ statistic}
The detection statistic that we use is described in \cite{schutz2}.
As in \cite{schutz2} we call this statistic $\calf$\footnote{
Note that this detection statistic has nothing to do with the F-statistic of the statistical literature, which is ratio of two sample variances, or the F-test of the null hypothesis that the two samples are drawn from distributions of the same variance.}, though
differences between our definition and that given in
\cite{schutz2} are pointed out below. 

The $\calf$ statistic derives from the method of maximum
likelihood. The log likelihood function, $\ln \Lambda$, is, for Gaussian noise
\beq
\label{e:likelihood}
\ln \Lambda = (s|h) - \frac{1}{2} (h|h) \;,
\eeq
where
\beq
\label{e:innerproduct}
(s|y) = 4 \Re \int _0^\infty \frac{\tilde{s}(f)\tilde{y}^\star(f)}
{S_n(f)} \rd f\;.
\eeq
$s$ is the calibrated detector output time series, $h$ is the target signal (commonly
referred to as the template), $\tilde{\phantom{y}}$~is the Fourier
transform operator, and $S_n(f)$ is the one-sided power spectral
density of the noise. The $\calf$ statistic is the maximum value
of $\ln \Lambda$ with respect to all the unknown signals parameters, given our data and a set of known template parameters. In fact, if some or all of the signal's parameters are unknown, it is
standard practice to compute the likelihood for different template
parameters and look for the highest values. The maximum of the
likelihood function is the statistic of choice for matched
filtering methods, and it is the optimal detection statistic as
defined by the Neyman-Pearson criterion: lowest false dismissal
rate at a fixed false alarm rate (see, for example,
Section~\ref{ss:sensitivity}).

In our case the known parameters are the position of
the source ($\alpha$, $\delta$ angles on the celestial sphere),
the emission frequency $f_{\rm s}$ and the first order spin-down parameter
value $\dot{f}_{\rm s}$. The unknown parameters are the orientation of
the pulsar (angle $\iota$), the polarization state of the wave
(angle $\psi$), its initial phase $\phi_0$, and the wave's
amplitude $h_0$.

The core of the calculation of $\calf$ consists in computing
integrals of the type given in Eq.~\ref{e:innerproduct}, using
templates for the two polarizations of the wave.  The results are
optimally combined as described in \cite{schutz2} except we
consider a single frequency component signal. Also, we do
not treat $S_n(f)$ as constant in time.  Thus, while the method is
defined in \cite{schutz2} in the context of stationary Gaussian
noise, we adapt it so that it can be used even when the noise is
nonstationary.  The calculation is easily performed in the
frequency domain since signal energy is concentrated in a narrow
frequency band. Using the SFTs described in \ref{sss:sfts} some
approximations can be made to simplify the calculation and improve
computational efficiency while still recovering most ($> 98\%$) of
the signal power.

The method of computing $\calf$ was developed for a specific
computational architecture: a cost-effective Beowulf cluster,
which is an ensemble of loosely-coupled processors with simple
network architecture.  This becomes crucial when exploring very
large parameter-space volumes for unknown sources using long
observation periods, because the search depth and breadth are
limited by computational resources.  The S1 analyses described
here were carried out using Condor \cite{condor} on the Merlin and
Medusa clusters at the AEI and UWM respectively
\cite{medusa,merlin}.  Each cluster has 300 independent CPUs.

As a point of reference we
note that it takes of order of a few seconds of CPU time on a 1.8
GHz-class CPU to determine the $\calf$ statistic for a single
template with $\sim$ 16\,d of observation time.

\subsubsection{Setting an upper limit on $h_0$}
The outcome ${\calf}^\star$ of a specific targeted search represents
the optimal detection statistic for that search. Over an independent
ensemble of similar searches in the presence of stationary Gaussian noise,
${2{\calf}^\star}$ is a random variable that follows a $\chi^2$ distribution
with four degrees of freedom. If the data also contain a signal, this
distribution has a
non-centrality parameter $\lambda$ proportional to the time-integral
of the squared signal.

Detection of that signal would be signified by a large value ${\calf}^\star$ unlikely to
have arisen from the noise-only distribution. If instead the value
is consistent with pure noise (as we find in this instance),
we can place an upper limit on the strength
of any signal present, as follows:

Let $\calf^\star$ be the value of the detection statistic in our actual experiment. Had there existed in the data a real signal with amplitude
greater than or equal to
$h_0(C)$, then in an ensemble of identical experiments with different realizations of the noise, some fraction of
trials $C$ would yield a detection statistic exceeding the value ${{\calf}^\star}$. We will therefore say that we have
placed an upper limit $h_0(C)$ on the strength of the targeted signal, with
confidence $C$. This is a standard frequentist
upper limit.

To determine the probability distribution $p(2{\calf}|h_0)$, we produce a set of simulated
artificial signals with fixed amplitude $h_0$ from fictional pulsars at the
position of our target source, and with the same spin-down parameter
value, but with intrinsic emission frequencies that differ from it by
a few tenths of a hertz.  We inject each of these artificial signals into
our data and run a search with a perfectly matched template. For each
artificial signal we obtain an independent value of the detection statistic;
we then histogram these values. If the SFT data in nearby frequency
bins (of order 100 bins) can be considered as different
realizations of the same random process (justified in \ref{ss:characterization}),
then it is reasonable to assume 
that the normalized histogram represents the probability density
function $p(2{\calf}|h_0)$.  One can then compute the
confidence
\beq
\label{E:FStatC} C(h_0)=\int_{2{\calf}^\star}^\infty~ p(2{\calf} \vert
h_0) \rd(2{\calf}),
\eeq
where $h_0(C)$ is the functional inverse of $C(h_0)$.
In practice, the value of the integral in Eq.~\ref{E:FStatC} is calculated directly from our
simulations as follows: we count how many values of $\calf$ are
greater or equal to ${\calf}^\star$ and divide this number by the
total number of $\calf$ values. The value
derived in this way does not rely on any assumptions about the shape
of the probability distribution function (pdf) curve $p(2{\calf}| h_0)$.

There is one more subtlety that must be addressed: all eight signal
parameters must be specified for each injected artificial signal. The values of source
position and spin-down parameters  are known from radio data and are used for
these injections. Every injected signal has a different frequency but
all such frequencies lie in bins that are
close to the expected frequency of the target signal,
$1\,283.86$\,Hz. The values $\iota$ and $\psi$
are not known, and no attempt has been made in this
analysis to give them informative priors based on radio data.  However, the
value of the non-centrality parameter that determines the $p(2{\calf}|h_0)$
distribution does depend on these values. This means that for a given
${\calf}^\star$, a different confidence level can be assigned for
the same signal strength, depending on the choice of $\iota$ and
$\psi$.

There are two ways to proceed: either inject a population
of signals with different
values of $\iota$ and $\psi$, distributed according to the priors
on these parameters\footnote{The time domain analysis assumes
uniform priors on $\cos\iota$ and $\psi$.}, or pick a single value
for $\iota$ and for $\psi$. In the latter case it is reasonable
to choose the most pessimistic orientation and
polarization of the pulsar with respect to the detector during the
observation time. For fixed signal strength, this choice results
in the lowest confidence level and thus, at fixed confidence, in the
most conservative upper limit on the signal strength.  We have
decided to use in our signal injection the worst-case values for
$\iota$ (which is always $\pi/2$) and $\psi$, i.e., the values for which the non-centrality
parameter is the smallest.

\subsubsection{The frequency domain S1 analysis for PSR J1939+2134}
Table \ref{t:fdAnalyses} summarizes the results of the frequency domain analysis.
For every interferometer (column 1) the value of the detection statistic
for the search for J1939+2134 is reported: $2{\calf}^\star$, shown in
column 4. Next to it is the corresponding value of the chance probability:
 \beq
 P_0(2{\calf}^\star)=\int_{2{\calf}^\star}^\infty ~p(2{\calf}\vert h_0=0) \rd(2\calf) ,
 \eeq
our estimate of how frequently one would expect to observe the
measured value of ${\calf}^\star$ or greater in the absence of a
signal. The values of $P_0(2{\calf}^\star)$ are not significant;
we therefore conclude that there is no evidence of a signal and
proceed to set an upper limit.

$T_{\rm{obs}}$ is the length of the live-observation time.
$h_0^{\rm{inject}}$ is the amplitude of the population of
injected signals that yielded a $95\%$ confidence.
The upper limit $h_0^{\rm{95\%}}$ differs from $h_0^{\rm{inject}}$
only by the calibration uncertainty, as explained in Section~\ref{ss:uncertainties}.
$C_{{\rm{meas}}}$ is the confidence level derived from the
injections of fake signals, and $\Delta C$ its estimated
uncertainty due to the finite sample size of the simulation.

The quantities in the remaining columns can be used to evaluate
how far the reported results are from those that one expects. The
results shown are remarkably consistent with what one expects
based on the noise and on the injected signal: the confidence
levels that we determine differ from the expected ones by less
than  $2\%$.

Given a perfectly matched template, the expected non-centrality parameter when a signal $h(t)$ is added to white noise with spectral density
$S_n$ is
 \beq
 \label{e:lambda1}
 \lambda = \frac{2U}{S_n },
 \eeq
where $U=\int_{T_{\rm obs}} |h(t)|^2 \rd t$. $U$ can also be
computed by feeding the analysis pipeline pure signal and by
performing the search with a perfectly matched template
\footnote{This is indeed one of the consistency checks that have
been performed to validate the analysis software. We have verified
that the two values of $U$ agree within a $1\%$ accuracy.} having
set $S_n(f)=1$\,s. In Table \ref{t:fdAnalyses} we report the
values of $U_0$, for the worst-case $h(t)$ signals for
PSR J1939+2134 as `seen' by the interferometers during their
respective observation times and with $h_0=2\times 10^{-19}$. The
different values of $U_0$ reflect the different durations of the
observations and the different orientations of each detector with
respect to the source. The expected value of the non-centrality
parameter can be estimated as:
 \beq
 \label{e:lambdaexp}
 \lambda_{\rm{exp}} = {2U_0}\langle 1/S_n \rangle ~\left(  \frac{h_0^{\rm{inject}}}
                                  {2\times 10^{-19}  }\right )^2 .
 \eeq
If the noise were stationary, then $S_n$ may be easily determined.
Our noise is not completely stationary, so the value determined
for the non-centrality parameter $\lambda$ is sensitive to the
details of how $S_n$ is estimated. The value of $\langle 1/S_n \rangle$ used to 
determine the expected value of $\lambda$ is computed as
 \beq \langle
 1/S_n\rangle  = \frac{\Delta t}{M} \sum_\alpha
                           \frac{1}{\sum^M_k |\tilde x_{\alpha k}|^2/M},
 \label{e:ShEstimate}
 \eeq
where the frequency index $k$ varies over a band $\sim 0.2$\,Hz
around $1\,283.89$\,Hz. $N$ and $\Delta t$ are the number of
samples and the sampling time of the 60\,s time series that are
Fourier transformed. We choose an harmonic mean rather than an
arithmetic mean because this is the way $S_n$ enters the actual
numerical calculation of the $\calf$-statistic. This method is
advantageous because the estimate it produces is relatively
insensitive to very large outliers that would otherwise bias the
estimate.

$\lambda_{{\rm{exp}}}$ is the expected value of the non-centrality
parameter based on $S_n$ and $h_0^{\rm{inject}}$.
$\lambda_{{\text{best-fit}}}$ is the best-fit value of the
non-centrality parameter based on the measured distribution of
$\calf$ values from the simulation. $C_{{\rm{exp}}}$ and
$C_{{\text{best-fit}}}$ are the confidence levels corresponding to
these distributions integrated between $2{\calf}^\star$ and
$\infty$.
\begin{widetext}

\begin{table}

\begin{center}
\begin{tabular}{c|c|c|c|c|c|c|c|c|c|c|c}
  \hline
  \hline
  IFO & $T_{\rm{obs}}$ [d]& $h_0^{{\rm{inject}}}$& $2{\calf}^\star$     & $P_0(2{\calf}^\star)$ &
$\langle 1/S_n\rangle^{-1}$ [Hz$^{-1}$] & $U_0/10^{-33}$ [s]&$\lambda_{{\rm{exp}}}$& $\lambda_{{\text{best-fit}}}$& $C_{{\rm{exp}}}$ & $C_{{\text{best-fit}}}$ & $C_{{\rm{meas}}}\pm\Delta C$\\
  \hline
  \hline
  GEO   & 16.7   &$1.94\times 10^{-21}$  & $1.5$ &0.83 & $5.3\times 10^{-38}$ & $1.0$ & $3.6$ & $3.3$ & $95.7\%$ & $95.2\%$ & $95.01\pm 0.23\%$ \\
  \hline
  L1  & 5.73    &$2.70\times 10^{-22}$  & $3.6$ &0.46 & $1.4\times 10^{-40}$ & $0.37$ & $9.6$ & $8.3$ & $96.7\%$ & $95.0\%$ & $95.00\pm 0.23\%$ \\
  \hline
  H1  & 8.73   & $5.37\times 10^{-22}$   & $6.0$ & 0.20 & $5.4\times 10^{-40}$ & $0.5$ &   $13.3$ & $12.8$ & $96.6\%$  & $95.0\%$ & $95.00\pm 0.23\%$  \\
  \hline
  H2  & 8.90  &$3.97\times 10^{-22}$  & $3.4$ &0.49 & $3.8\times 10^{-40}$ &  $0.45$ & $9.3$ & $7.9$ & $96.8\%$ & $95.0\%$ & $95.00\pm 0.23\%$ \\
  \hline
  \hline
\end{tabular}
\end{center}

\caption{Summary of the frequency domain search analyses. $T_{\rm
{obs}}$ indicates the total duration of the analyzed data set.
$\calf^\star$ is the measured value of the detection statistic.
$P_0(2\calf^\star)$ is the probability of getting this value or
greater by chance, i.e. in the absence of any signal.
$h_0^{\rm{inject}}$ is the amplitude of the population of fake
signals that were injected in the data set such that, when
searched for with a perfectly matched template, $C_{\rm {meas}}$\%
of the time the resulting value of $\calf$ was greater than
$\calf^\star$. $\langle 1/S_n\rangle$ is the average value of the
inverse of the noise in a small frequency band around the target
frequency. $U_0$ is the time integral of the square of the
targeted signal with an amplitude of $2\times 10^{-19}$, at the output of
the interferometers, for observations times equal to
$T_{\rm{obs}}$ and in the absence of noise. $\lambda_{\rm{exp}}$
is the value of the non-centrality parameter that one expects for
the distribution of $\calf$ from searches with perfectly matched
templates on a population of injected signals with amplitude
$h_0^{\rm{inject}}$ and noise with average power  $\langle
1/S_n\rangle^{-1}$. $\lambda_{\rm{best-fit}}$ is the best-fit
non-centrality parameter value derived from the distribution
$p(2{\calf}\vert h_0^{\rm{inject}} )$ derived from the software
signal injections and searches with perfectly matched templates.
$C_{\rm{exp}}$ and $C_{\rm{best-fit}}$ are the corresponding
confidence values for $\calf^\star$.}
 \label{t:fdAnalyses}
 \end{table}
 \end{widetext}

\begin{figure}
\centering \includegraphics[height=6.5cm]{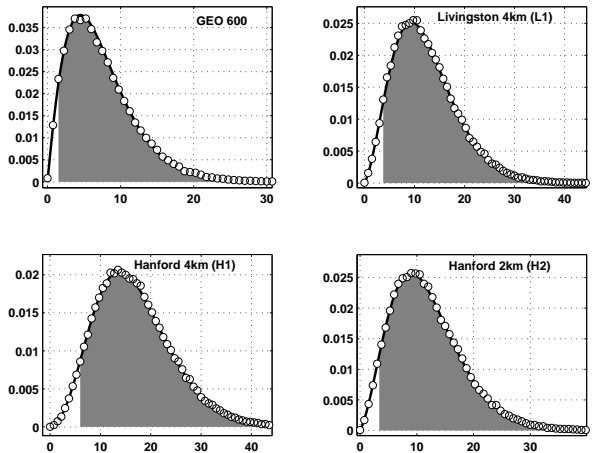} \caption{Measured
pdf for $2\calf$ for all four interferometer data
with injected signals as described in Table \ref{t:fdAnalyses}.
The circles represent the measured pdf values from the Monte Carlo
simulations.  The lines represent $\chi^2$ distributions with 4
degrees of freedom and best-fit non centrality parameters given in
Table \ref{t:fdAnalyses}. The filled area represents the integral
of the pdfs between $2{\calf}^\star$ and $+\infty$.}
\label{f:J1939FpdfA0.009}
\end{figure}

Fig.~\ref{f:J1939FpdfA0.009} shows the distributions for
$p(2{\calf} \vert h_0^{\rm{inject}})$. The circles result from the
simulations described above. The solid lines show the best fit
non-central $\chi^2$ curves. The shaded region is the integral of
$p(2{\calf} \vert h_0^{\rm{inject}})$ between  $2{\calf}^\star$
and $\infty$. By definition this area is 0.95.

\subsection{The time domain search technique}
\subsubsection{Overview}
\label{s:time}
Frequency domain methods offer high search efficiencies when the
frequency of the signal and/or the position of the neutron star
are unknown and need to be determined along with the other signal
parameters. However in the case of known pulsars, where both the
intrinsic rotation frequency of the neutron star and its position
are known to high accuracy, alternative time domain methods become
attractive. At some level the two domains are of course
equivalent, but issues such as data dropouts and the handling of
signals with complicated phase evolutions can be conceptually (and
practically) more straightforward in a time series analysis than
in an analysis based on Fourier transforms.

The time domain search technique employed here involves
multiplying (heterodyning) the quasi-sinusoidal signal from the
pulsar with a unit-amplitude complex function that has a phase
evolution equal but of opposite sign to that of the signal.  By
carefully modelling this expected phase, $\Phi(t)$, we can take
account of both the intrinsic frequency and spin-down rate of the
neutron star and its Doppler shift. In this way the
time-dependence of the signal is reduced to that of the strain
antenna pattern, and we are left with a relatively simple
model-fitting problem to infer the unknown pulsar parameters
$h_0$, $\iota$, $\psi$ and $\phi_0$ defined in Eqs~\ref{h} and
\ref{phase1}.

In the time domain analysis we take a Bayesian approach and
therefore express our results in terms of posterior
probability distribution functions for the parameters of
interest.  Such pdfs are conceptually very different from those used
to describe the $\cal{F}$ statistic used in the frequency domain
search, and represent the distribution of our \emph{degree of
belief} in the values of the unknown parameters, based on the
experiments and stated prior pdfs.

The time domain search algorithm comprises stages of heterodyning,
noise estimation and parameter estimation.  In outline, the data
are first heterodyned at a constant frequency close to the
expected frequency of the signal, low-pass filtered to suppress
contamination from strong signals elsewhere in the detector band
and re-binned to reduce the sampling frequency from 16\,384\,Hz to
4\,Hz.  A 
second (fine) heterodyne is applied to the data to account for the
time-varying Doppler shift and spin-down of the pulsar and any
final instrumental calibration, and the data are re-binned to
1~sample per minute. We take the data as stationary during this
period, and make an estimate of the noise variance in each 1 min
bin from the variance and covariance of the data contributing to
that bin. This variance is used in the likelihood function
described below.

The parameter estimation stage, at which we set the Bayesian upper
limit on $h_0$, proceeds from the joint probability of
these 1-min complex samples, $\{B_k\}$. We take these $B_k$ values
to have a Gaussian likelihood with respect to our signal model,
$y(t_k;\mathbf{a})$, where $\mathbf{a}$ is a vector in our
parameter space with components ($h_0,\iota,\psi,\phi_0)$ and
$t_k$ is the time-stamp of the $k$th sample. The signal model, the
complex heterodyne of Eq.~\ref{h}, is
 \begin{align}
 y(t_k;\mathbf{a}) & = \frac{1}{4}F_{+}(t_k;\psi)h_{0} (1 +
  \cos^{2}\iota)e^{i2\phi_{0}}\nonumber\\
  & - \frac{i}{2}F_{\times}(t_k;\psi) h_{0} \cos\iota \,e^{i2\phi_{0}}.
 \end{align}
We choose uniform prior probabilities for $\phi_0$ over $[0,2\pi]$
and $\psi$ over $[-\pi/4,\pi/4]$, and a prior for $\iota$ that is
uniform in $\cos\iota$ over $[-1,1]$, corresponding to a uniform
probability per unit solid angle of pulsar orientation. These
uniform priors are uninformative in the sense that they are
invariant under changes of origin for the parameters. Although
strictly a scale parameter, the prior for $h_0$ is also chosen as
constant for $h_0\geq 0$, and zero $h_0<0$. This is a highly
informative prior, in the sense that it states that the prior
probability that $h_0$ lies between $10^{-24}$ and $10^{-25}$ is
10 times less than the prior probability it lies between
$10^{-23}$ and $10^{-24}$, but guarantees that our posterior pdf
can be normalized.

The joint posterior pdf for these parameters is
\begin{align}
 p(\mathbf{a}|\{B_k\})\propto p(\mathbf{a}) & \exp\left[-\sum_{k}\frac{\Re \{ B_{k}-
 y(t_k;\mathbf{a}) \} ^2}{2\sigma_{\Re \{ B_k \}} ^2} \right] \nonumber \\
 & \times ~ \exp \left[ -\sum_{k}\frac{\Im \{ B_{k}- y(t_k;\mathbf{a}) \} ^2}{2\sigma_{\Im \{ B_k \} }^2}\right],
\end{align}

where $p(\mathbf{a})$ $(\propto \sin\iota)$ is the prior on
$\mathbf{a}$, $\sigma_{\Re \{ B_k  \} }^2$ is the variance of the real parts 
of $B_k$, and $\sigma_{\Im \{ B_k  \} }^2$ is the variance of the imaginary parts 
of $B_k$.

The final stage in the analysis is to integrate this posterior pdf
over the $\iota$, $\psi$ and $\phi_0$ parameters to give a
marginalized posterior for $h_0$ of
 \begin{equation}
 p(h_0|\{B_k\}) \propto \iiint p(\mathbf{a}|\{B_k\})\rd\iota\rd\psi\rd\phi_0,
 \end{equation}
normalized so that $\int_0^\infty p(h_0|\{B_k\})\rd h_0=1$. This
curve represents the distribution of our degree of belief in any
particular value of $h_0$, given the model of the pulsar signal,
our priors for the pulsar parameters, and the data. The width of
the curve roughly indicates the range in values consistent with
our state of knowledge.

By definition, given our data and priors, there is a probability
of 0.95 that the true value of $h_0$ lies below $h_0^{95\%}$ where
 \begin{equation}
 0.95 = \int_0^{h_0^{95\%}}p(h_0|\{B_k\})\rd h_0,
 \end{equation}
and this defines our 95\%-credible Bayesian upper limit on $h_0$.

An attraction of this analysis is that data from different
detectors can be combined directly using the appropriate signal
model for each.  The combined posterior distribution from all the
available interferometers comes naturally out of a Bayesian
analysis and, for independent observations, is simply the
(normalized) product of the contributing probability
distributions, i.e.,
 \begin{equation}
 p(\mathbf{a}| \textrm{all data})\propto
  p(\mathbf{a}| \textrm{GEO})\times
   p(\mathbf{a}| \textrm{H1})\times
    p(\mathbf{a}| \textrm{H2})\times
     p(\mathbf{a}| \textrm{L1}).
 \end{equation}
This posterior pdf embodies all we believe we know about the
values of the parameters, optimally combining the data from all
the interferometers in a coherent way. For interferometers with
very different sensitivities, this will closely approximate the
result from the most sensitive instrument. Again, we must
marginalize over $\iota$, $\psi$ and $\phi_0$ to obtain the posterior pdf
for $h_0$ alone.
We note that this is more than simply a combination of the
marginalized pdfs from the separate interferometers as the
coherence between the instruments is preserved, and it recognizes the different polarization
sensitivities of each.

Equipment timing errors discovered after S1 cautioned against a
coherent multi-interferometer analysis with this data set.
In principle we could assign a suitable prior for the resulting phase offsets and marginalize over them. However, the dominant position of
the Livingston 4km interferometer means that even a fully a coherent analysis would only improve our sensitivity by about 20\%, so we have not pursued this.  Fully coherent analyses will be
possible in future observing runs.

\subsubsection{The time-domain S1 analyses for PSR J1939+2134}
The time domain search used
contiguous data segments 300\,s or longer in duration.

\begin{figure}[!]
\centering \includegraphics[width=8cm]{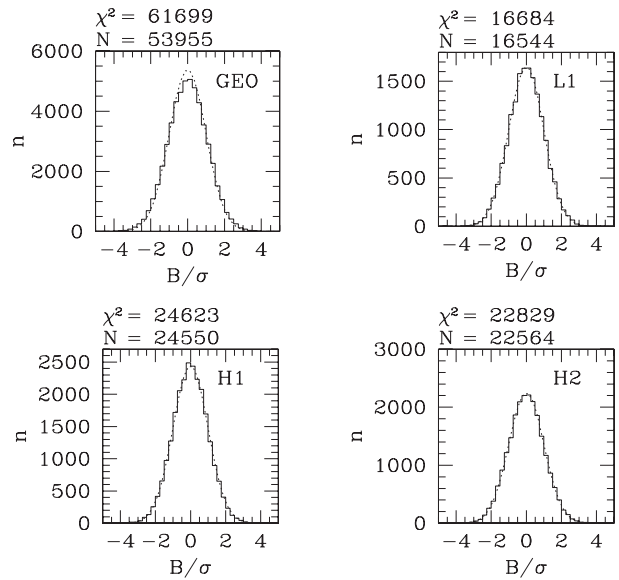}
\caption{Histograms of $B/\sigma = \Re(B_k)/\sigma_{\Re \{B_k  \}} +
\Im(B_k)/\sigma_{\Im \{ B_k \} }$ for each interferometer.  The dotted lines
represent the expected Gaussian distribution, with $\mu = 0$ and
$\sigma = 1$.} \label{histo_all}
\end{figure}

The effectiveness of the noise estimation procedure described
above was assessed from histograms of $B/\sigma = \Re(B_k)/\sigma_{\Re \{ B_k \}} +
\Im(B_k)/\sigma_{\Im \{ B_k  \} }$. If the estimates are
correct, and our likelihood function is well modelled by a
Gaussian, these histograms (Fig.~\ref{histo_all}) should also be
Gaussian with a variance of one.   Since we divide the noise
between the real and imaginary components we expect the value of
$\chi^2$ to be close (within $\sqrt{2N}$) of the number of real
and imaginary data,  $N$ (twice the number of complex binned data
values $B_k$). A small number of outliers with magnitudes of $B_k
/ \sigma_k$ larger than 5 were not included in this or subsequent
analyses.

\begin{figure}[!]
 \centering \includegraphics[height=8cm]{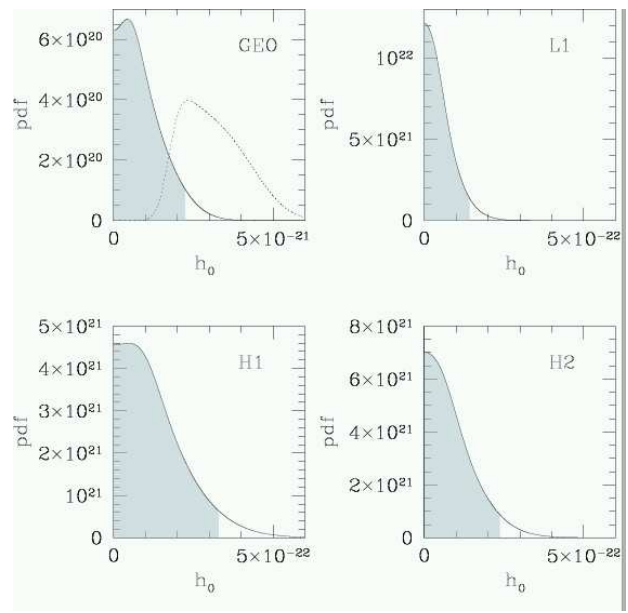}
 \caption{For each interferometer, the solid line represents the
marginalized posterior pdf for $h_0$ (PSR J1939+2134) resulting
from the S1 data. The 95\% upper limits (extent of the shaded
region) are $2.2 \times 10^{-21}$ for GEO, $1.4 \times 10^{-22}$
for L1, $3.3 \times 10^{-22}$ for H1 and $2.4 \times 10^{-22}$ for H2.  The dotted line in the GEO plot shows the posterior pdf
of $h_0$ in the presence of a simulated signal injected into the
GEO S1 data stream using $h_0 = 2.2\times10^{-21}$, $\phi_0 =
0^\circ$, $\psi = 0^\circ$ and $\iota = 0^\circ$. }
 \label{ligogeopdfs}
\end{figure}

The marginalized posterior pdfs for $h_0$ are plotted as the solid
lines in Fig.~\ref{ligogeopdfs}.  These represent the distribution
of our degree of belief in the value of $h_0$, following S1, for
each interferometer. The width of each curve roughly indicates the
range in values consistent with our priors and the data from the
instruments individually. The formal 95\% upper limits from this
analysis are the upper bounds to the shaded regions in the plots,
and are $2.2 \times 10^{-21}$ for GEO, $1.4
\times 10^{-22}$ for L1, $3.3 \times 10^{-22}$ for
H1, and $2.4 \times 10^{-22}$ for H2. 

The dotted line in the GEO plot of Fig.~\ref{ligogeopdfs} shows
the (very different) marginalized posterior pdf obtained when a
simulated signal is added to the data with an amplitude of
$2.2\times10^{-21}$, and with $\phi_0 = 0^\circ$, $\psi = 0^\circ$
and $\iota = 0^\circ$.  Here there is a clear non-zero lower limit
for the value of $h_0$, and a result such as this would have
indicated a nominal detection, had we seen it.

\subsection{Estimation of uncertainties}
\label{ss:uncertainties}

In the frequency domain analysis the uncertainty on the upper
limit value, $h_0^{95\%}$, has two contributions. The first stems
from the uncertainty on the confidence ($\Delta C\approx 0.23\%$)
that results from the finite sample size of the simulations. In
order to convert this uncertainty into an uncertainty on
$h_0^{95\%}$, we have performed several additional Monte Carlo
simulations. For every run we have injected a population of
signals with a given strength, $h_0^{\rm{inject}}$, near
$h_0^{95\%}$, searched for each of them with a perfectly matched
template and derived a value of ${\calf}$. With these values we
were able to estimate
the $h_0(C)$ curve near $h_0^{95\%}$, its slope $h_0^\prime$ and
from this the uncertainty on the value of $h_0^{\rm{inject}}$:
 \beq
 \Delta h_0^{95\%} \approx h_0^\prime \Delta C.
 \label{e:errorconf}
\eeq

The second contribution to the uncertainty on the value of
$h_0^{95\%}$ comes from errors in the calibration of the
instruments, which influence the absolute sensitivity scale. In
particular this reflects in an uncertainty in the actual value of
the strength of injected signals so that $h_0^{95\%} =
h_0^{\rm{inject}} \pm \delta h_0^{\rm {cal}}$. The sum of this
error, estimated in \ref{ss:configuration}, and the error arising
from the finite sample size, Eq.~\ref{e:errorconf}, is given in
the frequentist results in Table \ref{t:ULsummary}.

Note that when a pulsar signal is present in the data, errors in
the calibration introduce errors in the phase and amplitude of
that signal. The errors in $\calf$ {\em due to the signal} are
quadratic with the errors in the phase and are linear with the
errors in the amplitude. However the estimate of the noise
spectral density is also effected by calibration errors, and in
particular by the amplitude errors. The net effect on  $\calf$ is
that the resulting error on this quantity (which can be considered
a sort of signal-to-noise ratio) is quadratic in calibration
errors, thus insensitive, to first order, to calibration errors.

The errors quoted for the Bayesian results in Table
\ref{t:ULsummary} simply reflect the calibration uncertainties
given in \ref{ss:configuration}.  For clarity, no attempt has been
made to fold a prior for this calibration factor into the marginal
analysis.

\section{Conclusion}
\label{s:results}
\subsection{Summary of results}
Table \ref{t:ULsummary} summarizes the 95\%  upper limit (UL)
results that we have presented in the previous sections. We should
stress once more that the two analyses address two well-posed but
\emph{different} questions, and the common nomenclature is
somewhat misleading.

The frequentist upper limit statements made in Section
\ref{ss:frequency} refer to the likelihood of measuring a given
value of the detection statistic or greater in repeated
experiments, assuming a value for $h_0$ and a least-favorable
orientation for the pulsar. The Bayesian limits set in Section
\ref{s:time} refer to the cumulative probability \emph{of the
value of $h_0$ itself} given the data and prior beliefs in the
parameter values.  The Bayesian upper limits report intervals in which we are $95\%$ certain that the true value resides. We do not expect two such distinct definitions
of `upper limit' to yield the same numerical value.

Recall that the frequentist UL is \emph{conservative}: it is
calculated for the worst-case values of signal parameters $\iota$
and $\psi$ . The Bayesian TDS method marginalizes over these
parameters, gathering together the evidence supporting a
particular $h_0$ \emph{irrespective of orientation}. 
We have also performed an alternative calculation of the frequentist ULs by using a $p({\calf}|h_0)$ derived from 
a population of signals with $\cos\iota$ and $\psi$ parameters
uniformly distributed, as were the Bayesian priors in the time
domain search. As expected, we find that the resulting 
ULs have somewhat lower values than the conservative ones reported in Table~\ref{t:ULsummary}: $1.2\times 10^{-21}$, $1.5\times 10^{-22}$, $4.5\times 10^{-22}$ and $2.3\times 10^{-22}$ for the GEO, L1, H1 and H2 data sets respectively. 

\begin{table}
\begin{center}
\begin{tabular}{c|c|c}
  \hline
  IFO & Frequentist FDS       & Bayesian TDS \\
  \hline
  GEO   & $(1.9\pm 0.1)\times 10^{-21}$  & $(2.2\pm 0.1)\times 10^{-21}$  \\
  L1    & $(2.7\pm 0.3)\times 10^{-22}$  & $(1.4\pm 0.1)\times 10^{-22}$  \\
  H1    & $(5.4\pm 0.6)\times 10^{-22}$  & $(3.3\pm 0.3)\times 10^{-22}$  \\
  H2    & $(4.0\pm 0.5)\times 10^{-22}$  & $(2.4\pm 0.2)\times 10^{-22}$  \\
  \hline
\end{tabular}
\end{center}
\caption{Summary of the 95\% upper limit values of $h_0$ for PSR J1939+2134.
The frequency domain search (FDS) quotes a conservative frequentist upper limit and
the time domain search (TDS) a Bayesian upper limit after marginalizing over 
the unknown $\iota$, $\psi$ and $\phi_0$ parameters.}
\label{t:ULsummary}
\end{table}

\subsection{Discussion of previous upper limit results}
\label{s:s:previousUL}
Two prior upper limits have been published on the strain of a
signal from our specific  pulsar J1939+2134.  A limit of  $h < 3.1
\times 10^{-17}$ and $1.5 \times 10^{-17}$ for the first and
second harmonic of the rotation frequency of the pulsar,
respectively, was set in \cite{hereld83} using 4\,d of data from
the Caltech 40\,m interferometer. A tighter limit $h < 10^{-20}$
was determined using a divided-bar gravitational wave detector at
Glasgow University for the second harmonic alone \cite{hough83}.

More sensitive untargeted UL results on the strain of periodic GW
signals at other frequencies come from acoustic bar detector
experiments \cite{astone00}, \cite{astone01} and \cite{mauceli00}.
Due to the narrow sensitivity bands of these detectors (less than
1\,Hz around each mode), and the fact that their frequencies do
not correspond to those of any known pulsars\footnote{With the
exception of the Australian detector NIOBE and of the Japanese
torsional antenna built specifically to detect periodic signals
from the Crab pulsar \cite{suzuki95}.}, studies with bar antennas
have not investigated possible emission from any known pulsars.

In \cite{astone00} a UL of $2.9\times 10^{-24}$ was reported for
periodic signals from the Galactic center, with $921.32 < f_{\rm
s} < 921.38$\,Hz and no appreciable spin-down over $\sim 95.7$
days of observation. These data were collected by the EXPLORER
detector in 1991. This UL result was {\it not} obtained by a
coherent search over the entire observation time, due to
insufficient timing accuracy.

In \cite{astone01} a fully coherent 2 day-long all-sky search was
performed again on 1991 EXPLORER data in a $f_{\rm s}$ search band
of about 1\,Hz centered at 922\,Hz and including 1 spin-down
parameter. It resulted in an UL of $2.8\times 10^{-23}$ at the
$99\%$ confidence level. This search was based on the same
detection statistic used in our frequency domain analysis.

Another parameter space search is presented in \cite{mauceli00}.
Data taken from the ALLEGRO detector during the first three months
of 1994 were searched for periodic gravitational wave signals from
the Galactic center and from the globular cluster 47\,Tuc, with no resolvable spin-down
and with $f_{\rm s}$ in the two sensitive bands of their antenna,
$896.30-897.30$\,Hz and $919.76-920.76$\,Hz, with a 10\,$\mu$Hz
resolution. The resulting UL at $8\times 10^{-24}$ is reported.

There exist several results from searches using early broadband
interferometric detectors \cite{hough83, livas88, niebauer93,
jones95, zucker88,hereld83}. Due to the poor sensitivities of
these early detector prototypes, none of these upper limits is
competitive with the strain sensitivity achieved here. However,
many of the new issues and complications associated with broadband
search instruments were first confronted in these early papers,
laying the foundations for future analyses.

Data from the first science run of the TAMA detector were searched for continuous waves 
from SN1987A in a $0.05$ Hz band at $\sim 934.9$ Hz. 
The reported 99\% confidence upper limit was $ \sim 5 \times 10^{-23}$ \cite{tama03}.

Improved noise performance and longer observation times achieved with interferometric detectors
since S1 has made their sensitivities comparable to or better than
the narrow band peak sensitivity of the acoustic bars cited above,
over much broader bandwidths. Combined with the advances in
analysis methods presented in this paper we anticipate significant
advances in search depth and breadth in the next set of
observations.

\subsection{Upper limit on the ellipticity of the pulsar}
\label{s:epsilonUL}
An UL on $h_0$ for J1939+2134 can be interpreted as an UL on the
neutron star's equatorial ellipticity. Taking the distance to
J1939+2134 to be  $3.6$\,kpc, Eq.~\ref{h0_eps} gives an UL on
ellipticity corresponding to $h_0^{95\%} < 1.4\times 10^{-22}$ of
 \begin{equation}
\epsilon^{\rm{95\%}} = 2.9 \,\times 10^{-4}\: \left( \frac{10^{45} \,{\rm g ~ cm}^2}{I_{zz}}\right).
 \label{e:epsilon}
 \end{equation}

Of course, the UL on the ellipticity of J1939+2134 derived from S1
data is about five orders of magnitude higher than the UL obtained
from the pulsar's measured spindown rate: $\epsilon \leq 3.80 \times
10^{-9} (10^{45}{\rm \,g\, cm}^2/I_{zz})^{1/2}.$ However, an ellipticity
of $\sim 10^{-4}$ could in principle  be generated by an interior magnetic 
field of strength $\sim 10^{16}$\,G, or it could probably be
sustained in a NS with a solid core.  Therefore, the 
above exercise suggests that with improved detector
sensitivities, even a null result from a search
for {\em unknown} pulsars will place interesting constraints on
the ellipticities of rapidly-rotating neutron stars that might exist in our
galactic neighborhood.

\section{Acknowledgements}
The authors gratefully acknowledge the support of the United States National Science Foundation for the construction and operation of the LIGO Laboratory and the Particle Physics and Astronomy Research Council of the United Kingdom, the Max-Planck-Society and the State of Niedersachsen/Germany for support of the construction and operation of the GEO600 detector. The authors also gratefully acknowledge the support of the research by these agencies and by the Australian Research Council, the Natural Sciences and Engineering Research Council of Canada, the Council of Scientific and Industrial Research of India, the Department of Science and Technology of India, the Spanish Ministerio de Ciencia y Tecnologia, the John Simon Guggenheim Foundation, the David and Lucile Packard Foundation, the Research Corporation, and the Alfred P. Sloan Foundation.



%
\end{document}